\documentclass[aps,prd,showpacs]{revtex4-1}
\usepackage[latin1]{inputenc}
\usepackage{amsmath}
\usepackage{amsfonts}
\usepackage{amssymb}
\usepackage{amsthm}
\usepackage{bm}
\usepackage{mathrsfs}
\usepackage{tensor}
\usepackage{graphicx}

\begin{document}

\title{Dielectric Analog Space-Times}

\author{Robert T. Thompson}
\email{robert@cosmos.phy.tufts.edu}
\author{J\"{o}rg Frauendiener$^{\dag}$}
\affiliation{$^{*\,\dag}$ Department of Mathematics and Statistics,
University of Otago, P.O.\ Box 56, Dunedin, 9054,  New Zealand}
\affiliation{$^{\dag}$ Centre of Mathematics for Applications, University of Oslo, P.O.\ Box 1053, Blindern, NO-0316 Oslo, Norway}

\begin{abstract}
We generalize the notion of a dielectric analog Schwarzschild black hole model to analog models of arbitrary space-times; in particular, the approach is not restricted to static space-times.  This is done by establishing a correspondence between electrodynamics on a curved, vacuum manifold, with electrodynamics in a general linear dielectric residing in Minkowski space-time.  The mapping is not unique, allowing for some freedom in the specification of equivalent materials, which could be useful for exploiting recent developments in the production of metamaterials.   Some examples are considered, with special attention paid to the dielectric analog of the exterior Kerr geometry, which is found to be reproducible with regular, linear, dielectrics. 
\end{abstract}

\maketitle

\section{Introduction}
The idea of studying a gravitational system by replicating certain of its aspects in a laboratory environment through other, analogous, means has gained wide popularity today.  One of the main motivations behind the current interest in these analog systems is to study the Hawking effect, which shows that a black hole evaporates through a process of spontaneous particle creation and emission \cite{Hawking:1974sw}.  It was later shown that this type of particle creation is not unique to black hole space-times \cite{PhysRevD.14.870} but is a generic feature of any space-time possessing an apparent horizon \cite{Visser:2001kq}.

The process by which Hawking radiation is created is still not entirely understood, and seems virtually impossible to observe in gravitational systems.  The search for a non-gravitational physical system that possesses some kind of analog to the Hawking effect originated with so-called ``dumb holes'', comprising acoustic waves in moving fluids \cite{Unruh:1980cg,Unruh:1994je}, although a greater degree of success has been more recently achieved with surface waves, rather than acoustic waves, in moving fluids \cite{PhysRevD.66.044019,Rousseaux:2010md,Weinfurtner:2010nu}.  See Refs.\ \cite{lrr-2005-12, Novello:2002qg} for a review of other promising non-gravitational systems, including Bose-Einstein condensates and superfluids.  It is also possible to consider electromagnetic analog systems such as ``slow light'' propagating in highly dispersive media \cite{Leonhardt:2000fd,Novello:2002qg},  nonlinear electrodynamics \cite{PhysRevD.61.045001}, and light propagating in moving dielectric fluids \cite{DeLorenci:2002ws,Novello:2003}.  While a physical realization for electromagnetic analog space-times would pose challenges for reproducing the Hawking effect \cite{Schutzhold:2001fw,Unruh:2003ss}, it is -- in principle -- possible, and would certainly be useful for studying classical effects.  

A somewhat simpler electromagnetic approach is to identify the behavior of electromagnetic fields on a curved space-time manifold with the behavior of electromagnetic fields in a dielectric material residing in Minkowski space-time.  This idea seems to have been first advanced by Eddington to argue that the deflection of light passing through the curved space-time near a massive object would be indistinguishable from the deflection of light passing by the same object if the space-time were Minkowskian but filled with an appropriate dielectric material \cite{Eddington:1920}.  Gordon then examined the inverse problem of finding an equivalent curved space-time to describe a given dielectric material by an effective ``optical'' metric \cite{Gordon:1923}.  In a similar vein it was noticed by Plebanski that the form of the constitutive equations for electromagnetic fields in empty, curved, space-time is formally equivalent to the constitutive equations for electromagnetic fields in an appropriate dielectric \cite{Plebanski:1959ff}.  This formal equivalence has been exploited to study gravitational systems in only a few examples \cite{DeFelice:1971,Reznik:1997ag}. 

Two issues have contributed to the underdevelopment of this approach. One is that the materials required to mimic a gravitational system generally have somewhat unnatural characteristics, such as equal permeability and permittivity.  Another is that, as cautioned by Plebanski himself \cite{Plebanski:1959ff}, the Plebanski equations are not strictly covariant, which calls into question the limits of their applicability.  However, the recent development of man made ``metamaterials'' \cite{PhysRevLett.76.2480,Pendry:1999} allows for wide ranging control of material parameters, to the extent of even achieving a negative refractive index \cite{Smith:2004}.  Here we address the second issue by describing a completely covariant approach to dielectric analog space-times that generalizes the Plebanski approach.   Clearly there are close connections between metamaterials and analog space-times, and the method described here is closely related to a particular approach \cite{Thompson:2010a,Thompson:2010b} in ``transformation optics'', the emerging field of transformation-based metamaterial engineering \cite{Greenleaf:2003,Pendry:2006a,Leonhardt:2006a,Schurig:2006}.  While these are still the early days of metamaterial exploration, the approach has already shown the potential to address fundamental questions \cite{PhysRevLett.105.067402}.

This paper is organized as follows.  Section \ref{Sec:ClassicalEM} gives a brief review of electrodynamics in linear dielectric materials from a completely covariant, manifestly 4-dimensional standpoint.  Section \ref{Sec:Analogs} addresses the mapping of electromagnetic fields from a curved, vacuum, manifold to a material residing in Minkowski space-time.  The characteristic parameters describing the analog material are determined as a function of position.  Section \ref{Sec:Schwarzschild} reconsiders some examples known from the literature, not only demonstrating agreement with results obtained from the Plebanski equations but also illustrating some additional mappings allowed by this new approach. In Sec.\ \ref{Sec:WaveEqn} we examine more carefully the correspondence between null geodesics in the vacuum space-time and trajectories of light in material.  Section \ref{Sec:Kerr} considers the dielectric analog of the exterior Kerr geometry.  We conclude with Sec.\ \ref{Sec:Conclusions}.

\section{Classical Electrodynamics} \label{Sec:ClassicalEM}
The covariant description of electrodynamics in vacuum is thoroughly described in Ref.\ \cite{Misner:1974qy}, whose notation and sign convention we follow, as well as a multitude of other, very readable, sources, such as Ref.\ \cite{Baez}.  Here we briefly summarize a covariant description of electrodynamics in linear dielectric media that has been recently given \cite{Thompson:2010b}, but for a more complete study of electrodynamics in dielectric media see Ref.\ \cite{Hehl}.
 
Assume space-time to consist of a manifold $M$ that possesses a metric $\mathbf{g}$. The covariant approach to electrodynamics combines the electric field $\vec{E}$ and magnetic flux $\vec{B}$ into a single mathematical object, the field strength tensor $\mathbf{F}$, that in a locally Lorentz frame with Cartesian coordinates has components that can be represented as a matrix
\begin{equation} \label{Eq:FComponents}
 F_{\mu\nu} = \left(
 \begin{matrix}
  0 & -E_x & -E_y & -E_z\\
  E_x & 0 & B_z & -B_y\\
  E_y & -B_z & 0 & B_x\\
  E_z & B_y & -B_x & 0
 \end{matrix}
 \right).
\end{equation}
Additionally, the electric flux $\vec{D}$ and magnetic field $\vec{H}$ are combined into the excitation tensor $\mathbf{G}$, that can also be represented as a matrix
\begin{equation} \label{Eq:GComponents}
 G_{\mu\nu} = \left(
 \begin{matrix}
  0 & H_x & H_y & H_z\\
  -H_x & 0 & D_z & -D_y\\
  -H_y & -D_z & 0 & D_x\\
  -H_z & D_y & -D_x & 0
 \end{matrix}
 \right).
\end{equation}
The covariant Maxwell equations are expressed as
\begin{equation}
 \mathrm{d}\mathbf{F}=0, \quad \mathrm{d}\mathbf{G}=\mathbf{J},
\end{equation}
where $\mathrm{d}$ is the exterior derivative, and $\mathbf{J}$ is the charge-current 3-form.  

Furthermore, in a linear dielectric medium there exists a relationship between $\mathbf{F}$ and $\mathbf{G}$ given by the constitutive equation \cite{Thompson:2010a,Thompson:2010b}
\begin{equation} \label{Eq:Constitutive}
 \mathbf{G} = \boldsymbol{\chi}(\star\mathbf{F}),
\end{equation}
that in component form reads
\begin{equation} \label{Eq:ConstitutiveIndices}
 G_{\mu\nu} = \chi\indices{_{\mu\nu}^{\alpha\beta}}\star\indices{_{\alpha\beta}^{\sigma\rho}}F_{\sigma\rho}.
\end{equation}
In Eq.\ (\ref{Eq:Constitutive}), $\star$ is the Hodge dual on $(M,\mathbf{g})$, which for present purposes is to be understood as a map from 2-forms to 2-forms that has component form
\begin{equation} \label{Eq:star}
 \star\indices{_{\alpha\beta}^{\mu\nu}} = \frac12 \sqrt{|g|}\epsilon_{\alpha\beta\sigma\rho}g^{\sigma\mu}g^{\rho\nu}.
\end{equation}
The tensor $\boldsymbol{\chi}$ contains information on the dielectric material's properties (i.e.\ permittivity, permeability, and magneto-electric couplings), and can be thought of as representing an averaging over all the material contributions to an action that describes a more fundamental quantum field theory \cite{Hopfield:1958,Huttner:1992}.  We require $\boldsymbol{\chi}$ to be independently antisymmetric on its first two and last two indices, and in vacuum $\boldsymbol{\chi}(\star\mathbf{F}) = \star\mathbf{F}$.  This last condition means that the classical vacuum is treated as a linear dielectric with trivial $\boldsymbol{\chi}$, recovering the usual constitutive relations in vacuum.

The components of the constitutive equation provide a set of six independent equations that, in a local frame, can be collected in the form
\begin{equation} \label{Eq:ConstitutiveComponents1}
 H_a=(\check{\mu}^{-1})\indices{_a^b}B_b + (\check{\gamma_1}^*)\indices{_a^b}E_b, \  D_a=(\check{\varepsilon}^*)\indices{_a^b}E_b+ (\check{\gamma}_2^*)\indices{_a^b}B_b.
\end{equation}
where we use the notation $\check{a}$ to denote a $3\times 3$ matrix.  Rearranging these to 
\begin{equation} \label{Eq:ConstitutiveComponents2}
 B_a=(\check{\mu})\indices{_a^b}H_b + (\check{\gamma_1})\indices{_a^b}E_b, \  D_a=(\check{\varepsilon})\indices{_a^b}E_b+ (\check{\gamma_2})\indices{_a^b}H_b.
\end{equation}
gives a representation that may be more familiar. These three-dimensional representations of the completely covariant Eq.\ (\ref{Eq:Constitutive}) are essentially equivalent, and it is a simple matter to switch between them using the relations
\begin{equation} \label{Eq:ConstitutiveShift}
   \check{\varepsilon}=\check{\varepsilon}^*-\check{\gamma_2}^*\check{\mu}\check{\gamma_1}^*,\  \check{\gamma_1}=-\check{\mu}\check{\gamma_1}^*, \ \check{\gamma_2} = \check{\gamma_2}^*\check{\mu}.
\end{equation}
However, one should be aware that these $3\times 3$ matrices are not tensors but simply components of $\boldsymbol{\chi}$ that have been collected into matrices.  By matching with the usual representations above, one finds the matrix representation of $\boldsymbol{\chi}$ in a local orthonormal frame is \cite{Thompson:2010b}
\begin{equation} \label{Eq:CartesianChi}
 \chi\indices{_{\gamma\delta}^{\sigma\rho}} = \frac12\left(
\begin{matrix}
 \left(\begin{smallmatrix}
 0 & 0 & 0 & 0\\ 0 & 0 & 0 & 0\\ 0 & 0 & 0 & 0\\ 0 & 0 & 0 & 0\\
 \end{smallmatrix} \right) &

* &

* &

* \\[15pt]

 \left(\begin{smallmatrix}
 0 & -\mu^{-1}_{xx} & -\mu^{-1}_{xy} & -\mu^{-1}_{xz}\\ 
 \mu^{-1}_{xx} & 0 & -\gamma_{1xz} & \gamma_{1xy}\\ 
 \mu^{-1}_{xy} & \gamma_{1xz} & 0 & -\gamma_{1xx}\\ 
 \mu^{-1}_{xz} & -\gamma_{1xy} & \gamma_{1xx} & 0
 \end{smallmatrix} \right) &

 \left(\begin{smallmatrix}
 0 & 0 & 0 & 0\\ 0 & 0 & 0 & 0\\ 0 & 0 & 0 & 0\\ 0 & 0 & 0 & 0
 \end{smallmatrix} \right) &

* &

* \\[20pt]

 \left(\begin{smallmatrix}
 0 & -\mu^{-1}_{yx} & -\mu^{-1}_{yy} & -\mu^{-1}_{yz}\\ 
 \mu^{-1}_{yx} & 0 & -\gamma_{1yz} & \gamma_{1yy}\\ 
 \mu^{-1}_{yy} & \gamma_{1yz} & 0 & -\gamma_{1yx}\\ 
 \mu^{-1}_{yz} & -\gamma_{1yy} & \gamma_{1yx} & 0
 \end{smallmatrix} \right) &

 \left(\begin{smallmatrix}
 0 & -\gamma_{2zx} & -\gamma_{2zy} & -\gamma_{2zz}\\
 \gamma_{2zx} & 0 & -\epsilon_{zz} & \epsilon_{zy}\\
 \gamma_{2zy} & \epsilon_{zz} & 0 & -\epsilon_{zx}\\
 \gamma_{2zz} & -\epsilon_{zy} & \epsilon_{zx} & 0
 \end{smallmatrix} \right) &

 \left(\begin{smallmatrix}
 0 & 0 & 0 & 0\\ 0 & 0 & 0 & 0\\ 0 & 0 & 0 & 0\\ 0 & 0 & 0 & 0\\
 \end{smallmatrix} \right) &

* \\[20pt]

 \left(\begin{smallmatrix}
 0 & -\mu^{-1}_{zx} & -\mu^{-1}_{zy} & -\mu^{-1}_{zz}\\ 
 \mu^{-1}_{zx} & 0 & -\gamma_{1zz} & \gamma_{1zy}\\ 
 \mu^{-1}_{zy} & \gamma_{1zz} & 0 & -\gamma_{1zx}\\ 
 \mu^{-1}_{zz} & -\gamma_{1zy} & \gamma_{1zx} & 0
 \end{smallmatrix} \right) &

 \left(\begin{smallmatrix}
 0 & \gamma_{2yx} & \gamma_{2yy} & \gamma_{2yz}\\
 -\gamma_{2yx} & 0 & \epsilon_{yz} & -\epsilon_{yy}\\
 -\gamma_{2yy} & -\epsilon_{yz} & 0 & \epsilon_{yx}\\
 -\gamma_{2yz} & \epsilon_{yy} & -\epsilon_{yx} & 0
 \end{smallmatrix} \right) &

 \left(\begin{smallmatrix}
 0 & -\gamma_{2xx} & -\gamma_{2xy} & -\gamma_{2xz}\\
 \gamma_{2xx} & 0 & -\epsilon_{xz} & \epsilon_{xy}\\
 \gamma_{2xy} & \epsilon_{xz} & 0 & -\epsilon_{xx}\\
 \gamma_{2xz} & -\epsilon_{xy} & \epsilon_{xx} & 0
 \end{smallmatrix} \right) &

 \left(\begin{smallmatrix}
 0 & 0 & 0 & 0\\ 0 & 0 & 0 & 0\\ 0 & 0 & 0 & 0\\ 0 & 0 & 0 & 0
 \end{smallmatrix} \right) \\
\end{matrix} \right),
\end{equation}
where the $*$ indicates entries that are antisymmetric on either the first or second set of indices on $\chi\indices{_{\gamma\delta}^{\sigma\rho}}$.  Equation (\ref{Eq:CartesianChi}) represents $\boldsymbol{\chi}$ as a matrix of matrices; the first two indices of $\chi\indices{_{\gamma\delta}^{\sigma\rho}}$ give the $\gamma\delta$ component of the large matrix, which is itself a matrix described by the second set of indices on $\boldsymbol{\chi}$.

\section{Dielectric Analog Space-times} \label{Sec:Analogs}
The idea behind a dielectric analog space-time is to identify the behavior of electromagnetic fields on a curved space-time manifold with the behavior of electromagnetic fields in a dielectric material residing in Minkowski space-time.  More precisely, let $(\hat{M},\hat{\mathbf{g}})$ be a curved, vacuum, space-time manifold and let $\hat{\mathbf{F}}$ and $\hat{\mathbf{G}}$ be electromagnetic fields in the space-time.  In particular, it is assumed that $\hat{\mathbf{g}}$ is a solution of Einstein's equations, rather than the Einstein-Maxwell equations, so $\hat{\mathbf{g}}$ describes the background space-time onto which some electromagnetic fields have been placed.  Thus we want to identify the null geodesics of the curved, vacuum, space-time with trajectories through the equivalent dielectric medium as determined by the wave equation in that medium.  Because the curved space-time is vacuum, $\hat{\boldsymbol{\chi}}=\boldsymbol{\chi}_{vac}$.

\begin{figure}[h]
 \resizebox{\linewidth}{!}{\rotatebox{90}{\includegraphics{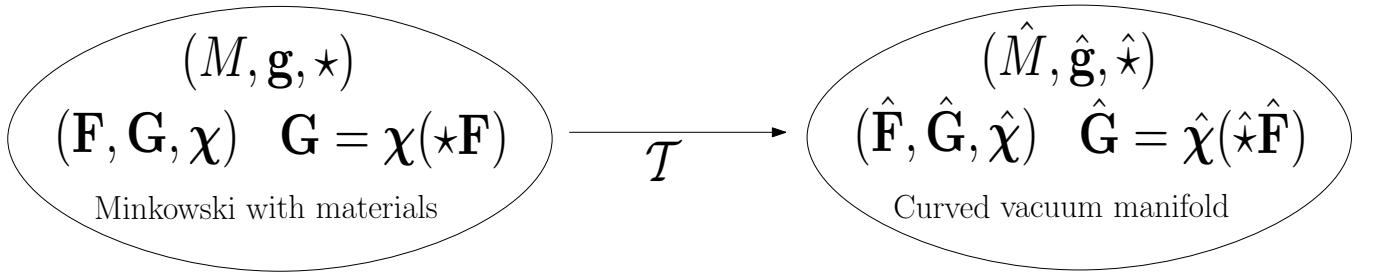}}}
 \caption[]{Electromagnetic fields $\hat{\mathbf{F}}$ and $\hat{\mathbf{G}}$ are pulled back from a patch of a curved, vacuum manifold to a material residing in Minkowski space-time by using the pullback $\mathcal{T}^*$.  \label{Fig:Map} } 
\end{figure}

The question that we would like to answer is: given a curved space-time, how do we determine the dielectric material that mimics this curved space-time?  The answer must come in the form of some characteristic material parameters such as permeability and permittivity.  To see how to obtain such material parameters as a function of position, consider a map $\mathcal{T}:M\to\hat{M}$ relating the electromagnetic fields in the material with electromagnetic fields in the curved space-time, as depicted in Fig.\ \ref{Fig:Map}.  

Because the metrics $\mathbf{g}$ and $\hat{\mathbf{g}}$ are known in each space-time, we demand that $\mathcal{T}$ acts only on the electromagnetic fields via its pullback, $\mathcal{T}^*$, by
\begin{equation}
 \mathcal{T}^*\left(\hat{\mathbf{F}}\right) = \mathbf{F}.
\end{equation}
Consider now the pulled-back fields.  At a point $x\in M$,
\begin{equation} \label{Eq:Gright}
 \mathbf{G}_x  = \mathcal{T}^* \left(\hat{\mathbf{G}}_{\mathcal{T}(x)}\right) 
 =  \mathcal{T}^*\left(\hat{\boldsymbol{\chi}}_{\mathcal{T}(x)} \circ \hat{\star}_{\mathcal{T}(x)}
 \circ \hat{\mathbf{F}}_{\mathcal{T}(x)}\right).
\end{equation}
But since $\mathbf{G}=\boldsymbol{\chi}(\star\mathbf{F})$ at $x\in M$, it is also true that
\begin{equation} \label{Eq:Gleft}
 \mathbf{G}_x=\boldsymbol{\chi}_x\circ \star_x \circ \, 
 \mathcal{T}^*\left(\hat{\mathbf{F}}_{\mathcal{T}(x)}\right).
\end{equation}
It follows that the right hand sides of Eqs.\ (\ref{Eq:Gright}) and (\ref{Eq:Gleft}) must be equal.  To clearly see what is going on, consider the action of $\mathbf{G}_x$ on a bi-vector $\mathbf{V}_x\in T^2_x(M)$, then
\begin{equation} \label{Eq:Grightleft}
 \left[\boldsymbol{\chi}_x\circ \star_x \circ \, 
 \mathcal{T}^*\left(\hat{\mathbf{F}}_{\mathcal{T}(x)}\right)\right](\mathbf{V}_x)=\hat{\boldsymbol{\chi}}_{\mathcal{T}(x)} \circ \hat{\star}_{\mathcal{T}(x)} \circ  
 \hat{\mathbf{F}}_{\mathcal{T}(x)} \circ \mathrm{d}\mathcal{T}(\mathbf{V}_x).
\end{equation}
Let $\Lambda\indices{^{\mu}_{\nu}}$ be the Jacobian matrix of $\mathcal{T}$, which is the matrix representation of $\mathrm{d}\mathcal{T}$.  Then Eq.\ (\ref{Eq:Grightleft}) can be written in component form as \cite{Thompson:2010a}
\begin{equation}
 \left(\chi\indices{_{\lambda\kappa}^{\xi\zeta}}\, 
 \star\indices{_{\xi\zeta}^{\gamma\delta}}\right)\Big|_x 
 \hat{F}_{\sigma\rho}\Big|_{\mathcal{T}(x)} \left(\Lambda\indices{^{\sigma}_{\gamma}} 
 \Lambda\indices{^{\rho}_{\delta}}\right)\Big|_x V^{\lambda\kappa}_x \\ 
 =\hat{\chi}\indices{_{\alpha\beta}^{\mu\nu}} \Big|_{\mathcal{T}(x)} 
 \hat{\star}\indices{_{\mu\nu}^{\sigma\rho}}\Big|_{\mathcal{T}(x)} 
 \hat{F}_{\sigma\rho}\Big|_{\mathcal{T}(x)} \left(\Lambda\indices{^{\alpha}_{\lambda}} 
 \Lambda\indices{^{\beta}_{\kappa}}\right)\Big|_x V^{\lambda\kappa}_x,
\end{equation}
where we have explicitly indicated where each object is evaluated. Eliminating $\hat{\mathbf{F}}$ and $\mathbf{V}_x$ from both sides and solving for $\boldsymbol{\chi}$ as a function of $x\in M$ gives
\begin{equation} 
 \chi\indices{_{\lambda\kappa}^{\tau\eta}}(x)=-\Lambda\indices{^{\alpha}_{\lambda}} \Lambda\indices{^{\beta}_{\kappa}} \hat{\chi}\indices{_{\alpha\beta}^{\mu\nu}}\Big|_{\mathcal{T}(x)} \hat{\star}\indices{_{\mu\nu}^{\sigma\rho}}\Big|_{\mathcal{T}(x)} (\Lambda^{-1})\indices{^{\pi}_{\sigma}}(\Lambda^{-1})\indices{^{\theta}_{\rho}}\, \star\indices{_{\pi\theta}^{\tau\eta}}\Big|_x.
\end{equation}
Here $\boldsymbol{\Lambda}^{-1}$ denotes the matrix inverse of $\boldsymbol{\Lambda}$, both $\boldsymbol{\Lambda}$ and $\boldsymbol{\Lambda}^{-1}$ are evaluated at $x$, and we have used the fact that on a 4-dimensional Lorentzian manifold, acting twice with $\star$ returns the negative, $\star\star\mathbf{F}=-\mathbf{F}$.
If the initial space-time is vacuum, then $\hat{\boldsymbol{\chi}}\hat{\star}=\hat{\star}$ and 
\begin{equation} \label{Eq:MaterialChi}
 \chi\indices{_{\lambda\kappa}^{\tau\eta}}(x)=-\Lambda\indices{^{\alpha}_{\lambda}} \Lambda\indices{^{\beta}_{\kappa}}  \hat{\star}\indices{_{\alpha\beta}^{\sigma\rho}}\Big|_{\mathcal{T}(x)} (\Lambda^{-1})\indices{^{\pi}_{\sigma}}(\Lambda^{-1})\indices{^{\theta}_{\rho}}\, \star\indices{_{\pi\theta}^{\tau\eta}}\Big|_x.
\end{equation}

Equation (\ref{Eq:MaterialChi}) gives the permeability, permittivity, and magneto-electric couplings of a material that mimics a curved, vacuum, manifold.  The metric $\hat{\mathbf{g}}$ of the curved space-time is arbitrary; in particular $\hat{\mathbf{g}}$ need not be static \cite{lrr-2005-12}.  If the map $\mathcal{T}$ is chosen to be the trivial coordinate identification $\mathcal{T}_0(\mathbf{x})=\mathbf{x}'$ (technically this map is not the identity, since the manifolds are different), then
\begin{equation} \label{Eq:IdMaterialChi}
 \chi\indices{_{\lambda\kappa}^{\tau\eta}}(x)=- \hat{\star}\indices{_{\lambda\kappa}^{\sigma\rho}}\Big|_{x} \star\indices{_{\sigma\rho}^{\tau\eta}}\Big|_x.
\end{equation}
This choice of $\mathcal{T}$ recovers results obtained with the Plebanski equations, but in general $\mathcal{T}$ need not be the trivial map $\mathcal{T}_0$, allowing for greater freedom in identifying materials.  For example, we show below how the exterior Schwarzschild geometry may be mapped into a spherical shell of finite thickness, albeit at the expense of introducing divergences in the material parameters.  It should be pointed out that the space-time represented by some dielectric configuration is not unique.  This is because the Hodge dual $\star$ is invariant under conformal transformation of the metric $\mathbf{g}\to\Omega^2\mathbf{g}$.  Therefore a dielectric analog space-time actually represents an equivalence class of conformally related space-times.

\section{Rindler Wedge and Exterior Schwarzschild Geometry} \label{Sec:Schwarzschild}

\subsection{Rindler Wedge}
As a simple example, consider the Rindler geometry described by the metric
\begin{equation}
 ds^2 =-a^2(x')^2(dt')^2+(dx')^2+(dy')^2+(dz')^2.
\end{equation}
The regions $x'<0$ and $x'>0$ correspond to the left and right Rindler wedges, respectively.  In this and what follows, primed coordinates denote the curved, vacuum manifold, while unprimed coordinates denote those of Minkowski space-time with dielectric media.  Thus the desired material parameters will be described as functions of unprimed coordinates. Choosing $\mathcal{T}=\mathcal{T}_0$, Eq.\ (\ref{Eq:IdMaterialChi}) returns a material described by scalar-valued permeability and permittivity
\begin{equation}
 \varepsilon=\mu=\frac{1}{ax}
\end{equation}
in agreement with results obtained from the Plebanski equations \cite{Reznik:1997ag}.  Notice that the material parameters diverge at the singular surface $x=0$.

\subsection{Exterior Schwarzschild Geometry}
The dielectric analog of spherically symmetric gravitational systems in isotropic coordinates was first studied by De Felice \cite{DeFelice:1971}, while Reznik \cite{Reznik:1997ag} mentions the material parameters representing the exterior Schwarzschild geometry.  Taking the line element in Schwarzschild coordinates 
\begin{equation}
 ds^2=-\left(1-\frac{2M}{r'}\right)(dt')^2+\left(1-\frac{2M}{r'}\right)^{-1}(dr')^2+(r')^2(d\theta')^2+(r')^2\sin^2\theta'(d\varphi')^2
\end{equation}
and the map $\mathcal{T}_0(t,r,\theta,\varphi)=(t',r',\theta',\varphi')=(t,r,\theta,\varphi)$ returns the material parameters
\begin{equation} \label{Eq:IdSchwarzschild}
 \varepsilon_{rr}=\mu_{rr}=1, \quad \varepsilon_{\theta\theta}=\varepsilon_{\varphi\varphi}=\mu_{\theta\theta}=\mu_{\varphi\varphi} = \left(1-\frac{2M}{r}\right)^{-1}
\end{equation}
in the representation of Eq.\ (\ref{Eq:ConstitutiveComponents2}).
Once again the horizon is characterized by divergent material parameters.  Also note that in both of the previous examples the material fills the entire space.  However, by adopting some map $\mathcal{T}\neq\mathcal{T}_0$, we may gain some freedom in specifying the boundaries of the material.  For example, adopting
\begin{equation}
 \mathcal{T}(t,r,\theta,\varphi)=(t',r',\theta',\varphi')=\left(t,\frac{2M(b-a)}{b-r},\theta,\varphi\right)
\end{equation}
maps the exterior Schwarzschild geometry onto a spherical shell $a\leq r\leq b$.  Using this map in Eq.\ (\ref{Eq:MaterialChi}) now leads to
\begin{equation}
 \varepsilon_{rr}=\mu_{rr}=\frac{2M(b-a)}{r^2}, \quad \varepsilon_{\theta\theta}=\varepsilon_{\varphi\varphi}=\mu_{\theta\theta}=\mu_{\varphi\varphi} = \frac{2M(b-a)^2}{(r-a)(b-r)^2}
\end{equation}
In a sense this result is worse in terms of the material, as its permeability and permittivity now diverge at both the inner and outer surfaces of the shell.  However, this behavior is not unexpected, and is similar to the example of compactifying a flat space-time containing an electric charge, which results in the appearance of an image charge on the boundary. 

We may go a step further and try to remove the divergences in $\check{\varepsilon}=\check{\mu}$ for the compactified exterior Schwarzschild analog by modulating the time coordinate with a map such as
\begin{equation}
  \mathcal{T}(t,r,\theta,\varphi)=(t',r',\theta',\varphi')=\left(\frac{2Mt(b-a)^2}{(r-a)(b-r)^2},\frac{2M(b-a)}{b-r},\theta,\varphi\right).
\end{equation}
We expect that this kind of map should eliminate the infinite blue shift by redefining time as the wavelength decreases towards the singular surfaces. Turning the crank on Eq.\ (\ref{Eq:MaterialChi}) returns
\begin{equation}
 \varepsilon_{rr}=\mu_{rr}=\frac{(r-a)(b-r)^2}{(b-a)r^2}, \quad \varepsilon_{\theta\theta}=\varepsilon_{\varphi\varphi}=\mu_{\theta\theta}=\mu_{\varphi\varphi} = 1,
\end{equation}
which is finite, as desired.  However, this type of spatially dependent time transformation generically introduces magneto-electric couplings \cite{Thompson:2010a}, which in this case turn out to be
\begin{equation}
 \gamma^1_{\theta\varphi}=-\gamma^2_{\theta\varphi}=\frac{(3r-2a-b)t}{(r-a)(b-r)}.
\end{equation}
Thus the divergences have been shifted from the permeability and permittivity to divergences in new magneto-electric couplings.  In the black hole space-time, ingoing light rays will be lost behind the horizon, so it is not surprising that trying to mimic this behavior in a material should require these essential singularities to appear somewhere. Note that not only do the magneto-electric couplings diverge at the inner and outer surface of the material shell, but there is also a sign change when $\check{\gamma_1}$, and $\check{\gamma}_2$ momentarily vanish at $r=\tfrac{2a+b}{3}$.  On the Schwarzschild manifold this corresponds to the point $r'=3M$, which is precisely the radius of the photon sphere in the Schwarzschild geometry.

\section{Wave Equation and Geometric Optics in Materials} \label{Sec:WaveEqn}
Before considering any further examples, return first to the constitutive equations and material tensor $\boldsymbol{\chi}$.  As they stand, Eqs.\ (\ref{Eq:ConstitutiveComponents1}) and (\ref{Eq:ConstitutiveComponents2}) are somewhat misleading, because while $E_a$ and $H_a$ are components of a 1-form, $D_a$ and $B_a$ are really selected components of the 2-forms $D_{ab}dx^a\wedge dx^b$ and $B_{ab}dx^a\wedge dx^b$.  Thus, a more appropriate version of Eq.\ (\ref{Eq:ConstitutiveComponents1}) would be something like
\begin{subequations} \label{Eq:ConstitutiveComponents3}
 \begin{equation}
  (\boldsymbol{\star}_{\Sigma}B)_a=(\check{\mu})\indices{_a^b}H_b + (\check{\gamma_1})\indices{_a^b}E_b,
 \end{equation}
 \begin{equation}
  (\boldsymbol{\star}_{\Sigma}D)_a=(\check{\varepsilon})\indices{_a^b}E_b+ (\check{\gamma_2})\indices{_a^b}H_b,
 \end{equation}
\end{subequations}
where $\star_{\Sigma}$ is the Hodge dual on an appropriate 3-dimensional spatial hypersurface.  But this requires that we resolve the space-time into space and time components, selecting an observer to define a direction of time.  The spatial hypersurface is then orthogonal to the selected direction of time.  Transforming to the local frame of the selected observer we can make the identifications of Eq.\ (\ref{Eq:CartesianChi}) and then give the constitutive equations a 3-dimensional representation.

In the Schwarzschild example above we expressed the results in spherical coordinates in the representation of Eq.\ (\ref{Eq:ConstitutiveComponents2}), but in dealing with the wave equation and determining the trajectories of light through a material it is the covariant $\boldsymbol{\chi}$ that is of real importance.
To see this, return again to Maxwell's inhomogeneous equation in the absence of sources
\begin{equation} \label{Eq:SourceFreeMaxwell}
 \mathrm{d}\mathbf{G}=0.
\end{equation}
Writing the exterior derivative of a 2-form in terms of the codifferential $\delta=-\boldsymbol{\star}\mathrm{d}\boldsymbol{\star}$, and using both the constitutive relation Eq.\ (\ref{Eq:Constitutive}) and the definition $\mathbf{F}=\mathrm{d}\mathbf{A}$, where $\mathbf{A}$ is the 4-potential, it follows that
\begin{equation} \label{Eq:WaveEqn}
 \delta\left(\boldsymbol{\star}\boldsymbol{\chi}\boldsymbol{\star}\mathrm{d}\mathbf{A}\right)=0
\end{equation}
is the covariant wave equation for the 4-potential \cite{Post:1962}.  We may reassure ourselves that since  $\boldsymbol{\chi}_{vac}\boldsymbol{\star}=\boldsymbol{\star}$ in vacuum, we recover the usual vacuum wave equation in the Lorenz gauge $\delta\mathbf{A}=0$,
\begin{equation}
 \delta\mathrm{d}\mathbf{A}=\triangle \mathbf{A} =0,
\end{equation}
where $\Delta$ is the Laplace-De Rham operator.  To see that the wave equation in the vacuum Schwarzschild space-time is equivalent to the wave equation in the analog material residing in Minkowski space-time, let us be more explicit with the notation and denote $\boldsymbol{\star}_{S}$ and $\boldsymbol{\star}_M$ as the Hodge dual in the Schwarzschild and Minkowski space-times respectively.  For illustrative purposes let us further assume the map $\mathcal{T}_0$.  Then from Eqs.\ (\ref{Eq:SourceFreeMaxwell}) and (\ref{Eq:IdMaterialChi}) it follows that
\begin{equation}
 \mathrm{d}(\boldsymbol{\chi}\boldsymbol{\star}_M\hat{\mathbf{F}}) = \mathrm{d}(\boldsymbol{\chi}_{vac}\boldsymbol{\star}_{S}\mathbf{F}),
\end{equation}
demonstrating the formal equivalence of the analog model on the level of the wave equation.  Thus, it is sufficient to describe the components of $\boldsymbol{\chi}$, which may then be transformed to the local frame of any observer.

We may go a step further in examining the covariant wave equation, and demonstrate that in the limit of geometric optics the trajectories of light rays in the equivalent analog media are equivalent to the null geodesics of the vacuum space-time (see also e.g., Refs.\ \cite{Post:1962,DeFelice:1971}). For this purpose let us assume the geometric optics ansatz for $\mathbf{A}$,
\begin{equation} \label{Eq:GeomOpAnsatz}
 A_{\mu}=\hat{A}_{\mu}e^{\frac{i}{\epsilon}k_{\nu}x^{\nu}},
\end{equation}
where $\epsilon$ is some dimensionless parameter.  Using this ansatz, Eq.\ (\ref{Eq:SourceFreeMaxwell}) may be calculated explicitly, and the geometric optics limit is obtained by taking $\epsilon\to 0$ \cite{Evans:1998}. We do this for the analog exterior Schwarzschild geometry described by Eq.\ (\ref{Eq:IdSchwarzschild}).  Consider first those modes with $k_{\mu}=(k_t,k_r,0,0)$, for which Eq.\ (\ref{Eq:SourceFreeMaxwell}) returns
\begin{multline}
  \left[k_t^2-\left(1-\frac{2M}{r}\right)^2k_r^2\right]\left(1-\frac{2M}{r}\right)^{-1}\sin\theta\left[\frac{\hat{A}_{\varphi}}{\sin^2\theta}\, \mathrm{d}t\wedge\mathrm{d}r\wedge\mathrm{d}\theta - \hat{A}_{\theta}\, \mathrm{d}t\wedge\mathrm{d}r\wedge\mathrm{d}\varphi\right] \\ +r^2\sin^2\theta(k_t\hat{A}_r-k_r\hat{A}_t)\left(k_t\,  \mathrm{d}t\wedge\mathrm{d}\theta\wedge\mathrm{d}\varphi + k_r\,  \mathrm{d}r\wedge\mathrm{d}\theta\wedge\mathrm{d}\varphi\right) = 0.
\end{multline}
Thus we find the condition on $k_{\mu}$ that
\begin{equation}
 k_t=\left(1-\frac{2M}{r}\right)k_r,
\end{equation}
which is precisely the condition determined by the requirement that $k_{\mu}k^{\mu}=0$ for radial null geodesics in the vacuum Schwarzschild space-time.  For a more general $k_{\mu}=(k_t,k_r,k_{\theta},k_{\varphi})$, calculating Eq.\ (\ref{Eq:SourceFreeMaxwell}) results in a more complicated expression that can be written as a system of four equations
\begin{subequations}
 \begin{equation}
  k_{\varphi}\left[-\hat{A}_tk_t\left(1-\tfrac{2M}{r}\right)^{-1}+\hat{A}_rk_r\left(1-\tfrac{2M}{r}\right)+\frac{\hat{A}_{\theta}k_{\theta}}{r^2}\right] \\ -\hat{A}_{\varphi}\left[-k_t^2\left(1-\tfrac{2M}{r}\right)^{-1}+k_r^2\left(1-\tfrac{2M}{r}\right)+\frac{k_{\theta}^2}{r^2}\right]=0,
 \end{equation}
 \begin{equation}
  k_{\theta}\left[-\hat{A}_tk_t\left(1-\tfrac{2M}{r}\right)^{-1}+\hat{A}_rk_r\left(1-\tfrac{2M}{r}\right)+\frac{\hat{A}_{\varphi}k_{\varphi}}{r^2\sin^2\theta}\right] \\ -\hat{A}_{\theta}\left[-k_t^2\left(1-\tfrac{2M}{r}\right)^{-1}+k_r^2\left(1-\tfrac{2M}{r}\right)+\frac{k_{\varphi}^2}{r^2\sin^2\theta}\right]=0,
 \end{equation}
 \begin{equation}
  k_{r}\left[-\hat{A}_tk_t\left(1-\tfrac{2M}{r}\right)^{-1}+\frac{\hat{A}_{\theta}k_{\theta}}{r^2}+\frac{\hat{A}_{\varphi}k_{\varphi}}{r^2\sin^2\theta}\right] \\ -\hat{A}_{r}\left[-k_t^2\left(1-\tfrac{2M}{r}\right)^{-1}+\frac{k_{\theta}^2}{r^2}+\frac{k_{\varphi}^2}{r^2\sin^2\theta}\right]=0,
 \end{equation}
 \begin{equation}
  k_{t}\left[\hat{A}_rk_r\left(1-\tfrac{2M}{r}\right)+\frac{\hat{A}_{\theta}k_{\theta}}{r^2}+\frac{\hat{A}_{\varphi}k_{\varphi}}{r^2\sin^2\theta}\right]-\hat{A}_{t}\left[k_r^2\left(1-\tfrac{2M}{r}\right)+\frac{k_{\theta}^2}{r^2}+\frac{k_{\varphi}^2}{r^2\sin^2\theta}\right]=0,
 \end{equation}
\end{subequations}
the simultaneous solution of which requires
\begin{equation}
 -\left(1-\frac{2M}{r}\right)^{-1}k_t^2+\left(1-\frac{2M}{r}\right)k_r^2+\frac{1}{r^2}k_{\theta}^2+\frac{1}{r^2\sin^2\theta}k_{\varphi}^2 = 0,
\end{equation}
corresponding to the condition $k^{\mu}k_{\mu}=0$ for a general null geodesic in the Schwarzschild space-time.  We have thus demonstrated that in the limit of geometric optics, the dielectric analog Schwarzschild black hole accurately mimics the behavior of light propagation in the vacuum space-time.

\section{Exterior Kerr Geometry} \label{Sec:Kerr}
Consider next the exterior Kerr geometry in Boyer-Lindquist coordinates described by the line element
\begin{equation}
  ds^2=-\left(1-\frac{2Mr'}{\rho^2}\right)(dt')^2 - \frac{4Mar'}{\rho^2}\sin^2\theta' dt'd\varphi' + \frac{\rho^2}{\Delta}(dr')^2+ \rho^2(d\theta')^2+\frac{\Sigma}{\rho^2}\sin^2\theta'(d\varphi')^2,
\end{equation}
where
\begin{equation}
 \rho^2=(r')^2+a^2\cos^2\theta', \quad \Delta=(r')^2-2Mr'+a^2, \quad
 \Sigma=\left((r')^2+a^2\right)^2-a^2\Delta\sin^2\theta',
\end{equation}
and $a$ is the ratio of angular momentum to mass, so that $L=aM$.  A naive calculation using the Plebanski equations \cite{Plebanski:1959ff,DeFelice:1971}, or using Eq.\ (\ref{Eq:IdMaterialChi}) and the representation Eq.\ (\ref{Eq:ConstitutiveComponents2}), returns material parameters like
\begin{equation}
 \varepsilon_{rr}=\mu_{rr}\propto \frac{\Delta}{\rho^2-2Mr},
\end{equation}
which diverge on the time-like limit surface (TLS), the boundary of the ergoregion.  This is a curious result.  The Boyer-Lindquist coordinates are well behaved at the TLS, so we should not expect divergent behavior at the corresponding material points.  However, in the previous section we have seen that $\boldsymbol{\chi}$ is really the only meaningful covariant quantity, and the map $\mathcal{T}_0$ provides the analog Kerr parameter in spherical coordinates as
\begin{equation} \label{Eq:KerrChi}
 \chi\indices{_{\gamma\delta}^{\sigma\rho}} = \frac12\left(
\begin{matrix}
 \left(\begin{smallmatrix}
 0 & 0 & 0 & 0\\ 0 & 0 & 0 & 0\\ 0 & 0 & 0 & 0\\ 0 & 0 & 0 & 0\\
 \end{smallmatrix} \right) &

* &

* &

* \\[15pt]

 \left(\begin{smallmatrix}
 0 & * & 0 & 0\\ \frac{r^2(\rho^2-2Mr)}{2\rho^2\Delta} & 0 & 0 & *\\ 0 & 0 & 0 & 0\\ 0 & \frac{aMr}{\rho^2\Delta} & 0 & 0\\
 \end{smallmatrix} \right) &

 \left(\begin{smallmatrix}
 0 & 0 & 0 & 0\\ 0 & 0 & 0 & 0\\ 0 & 0 & 0 & 0\\ 0 & 0 & 0 & 0
 \end{smallmatrix} \right) &

* &

* \\[20pt]

 \left(\begin{smallmatrix}
 0 & 0 & * & 0\\ 0 & 0 & 0 & 0\\ \frac{\rho^2-2Mr}{2\rho^2} & 0 & 0 & *\\ 0 & 0 & \frac{aM}{r\rho^2} & 0\\
 \end{smallmatrix} \right) &

 \left(\begin{smallmatrix}
 0 & 0 & 0 & 0\\ 0 & 0 & * & 0\\ 0 & \frac{\rho^2}{2\Delta} & 0 & 0\\ 0 & 0 & 0 & 0\\
 \end{smallmatrix} \right) &

 \left(\begin{smallmatrix}
 0 & 0 & 0 & 0\\ 0 & 0 & 0 & 0\\ 0 & 0 & 0 & 0\\ 0 & 0 & 0 & 0\\
 \end{smallmatrix} \right) &

* \\[20pt]

 \left(\begin{smallmatrix}
 0 & 0 & 0 & *\\ 0 & 0 & 0 & 0\\ 0 & 0 & 0 & 0\\ \frac{\Delta}{2\rho^2} & 0 & 0 & 0\\
 \end{smallmatrix} \right) &

 \left(\begin{smallmatrix}
 0 & * & 0 & 0\\ -\frac{aMr^3\sin^2\theta}{\rho^2\Delta} & 0 & 0 & *\\ 0 & 0 & 0 & 0\\ 0 & \frac{\Sigma}{2\rho^2\Delta} & 0 & 0\\
 \end{smallmatrix} \right) &

 \left(\begin{smallmatrix}
 0 & 0 & * & 0\\ 0 & 0 & 0 & 0\\ -\frac{aMr\sin^2\theta}{\rho^2} & 0 & 0 & *\\ 0 & 0 & \frac{\Sigma}{2r^2\rho^2} & 0\\
 \end{smallmatrix} \right) &

 \left(\begin{smallmatrix}
 0 & 0 & 0 & 0\\ 0 & 0 & 0 & 0\\ 0 & 0 & 0 & 0\\ 0 & 0 & 0 & 0
 \end{smallmatrix} \right) \\
\end{matrix} \right),
\end{equation}
where a $*$ denotes entries that are antisymmetric on either the first or second pair of indices of $\chi\indices{_{\gamma\delta}^{\sigma\rho}}$.
It is clear that $\boldsymbol{\chi}$ is completely regular and finite across the TLS, but that some of the components of $\boldsymbol{\chi}$ change sign at the TLS.  When the inverse of these components are used in the representation of Eq.\ (\ref{Eq:ConstitutiveComponents2}) a divergence appears.  In terms of the propagation of light, which depends on $\boldsymbol{\chi}$, everything is well behaved at the TLS.  The only divergences occur on the horizon, $\Delta=0$, as expected.

\section{Conclusions} \label{Sec:Conclusions}
Using a manifestly 4-dimensional covariant description of electrodynamics, a natural and covariant method of identifying a dielectric analog of a curved space-time has been developed.  First, the quantities that describe macroscopic electrodynamics in materials were encoded in such a way that the vacuum may be considered as just another dielectric material.  Then, by pulling back the fields from the curved space-time to Minkowski space-time with a material, the requisite material parameters could be solved for.  However, it was found that the map $\mathcal{T}$, whose pullback $\mathcal{T}^*$ operates on the fields, is not uniquely defined.

The non-uniqueness of $\mathcal{T}$ is somewhat unsettling, and it would be nice if there were a natural choice for $\mathcal{T}$.  On the other hand, the greater freedom in selecting $\mathcal{T}$ allows for greater manipulation of the resultant material, where for example, it is possible to map the exterior Schwarzschild geometry to a finite spherical shell residing in Minkowski space-time.  Such non-trivial mappings come at a price, such as more complicated material parameters, extra divergences, and induced magneto-electric couplings.

In analyzing the analog exterior Kerr geometry, it was found that the ``usual'' representation of the constitutive relations requires careful consideration as it may lead to spurious divergences.  However, the quantity with real relevance for the behavior of light in the material is $\boldsymbol{\chi}$ and the spurious divergences of the usual, 3-dimensional representation occur because they rely on the inverse of elements of $\boldsymbol{\chi}$ that may vanish at a point.

%

\end{document}